\documentclass[12pt]{article}
\begin{document}
\begin{center}

{\bf CROSSING OF THE w= -1 BARRIER IN VISCOUS MODIFIED GRAVITY }

\vspace{1cm}

Iver Brevik\footnote{Email: iver.h.brevik@ntnu.no}

\bigskip

Department of Energy and Process Engineering, Norwegian University
of Science and Technology, N-7491 Trondheim, Norway

\bigskip

\today
\end{center}

\begin{abstract}
We consider a modified form of gravity in which the action
contains a power $\alpha$ of the scalar curvature. It is shown how
the presence of a bulk viscosity in a spatially flat universe may
drive the cosmic fluid into the phantom region ($w<-1$) and thus
into a Big Rip singularity, even if it lies in the quintessence
region ($w>-1$) in the non-viscous case. The condition for this to
occur is that the bulk viscosity contains the power $(2\alpha-1)$
of the scalar expansion. Two specific examples are discussed in
detail. The present paper is a generalization of the recent
investigation dealing with barrier crossing in Einstein's gravity:
I. Brevik and O. Gorbunova, {\it Gen. Relativ. Grav.} {\bf 37},
2039 (2005).

\end{abstract}

\section{Introduction}

Modified versions of Einstein's gravity continue to attract
interest. The motivation for this kind of generalization has its
root in the well known observations of redshifts of  type Ia
supernovae \cite{reiss98,perlmutter99,tony03}, as well as the
anisotropy of the microwave background
\cite{bennet03,netterfield02,halversen02}. The data may be
explained by dark energy which in turn may result from a
cosmological constant, a cosmic fluid with a complicated equation
of state, a scalar field with  quintessence or phantom-like
behaviour, or perhaps by some other kind of theory. For a review
of the developments up to 2003, see Ref.~\cite{padmanabhan03};
some more recent papers are
Refs.~\cite{cognola06,sola06,capozziello06}. A forthcoming
extensive review is \cite{copeland06}.

The simplest way of dealing with the  expansion of the universe
mathematically, is to allow for a cosmological constant. It
corresponds to putting the parameter $w$ equal to -1 in the
equation of state for the cosmic fluid,
\begin{equation}
p=(\gamma -1)\rho  \equiv w\rho. \label{1}
\end{equation}
The fluid is then the same as the extreme tensile stress vacuum
"fluid" in the de Sitter universe. More general types of fluid can
be envisaged: thus the region $-1<w<-1/3$ corresponds to a
quintessence fluid, whereas the region $ w<-1$ corresponds to the
so-called phantom fluid.  It should be noted that in both these
cases $\rho +3p \leq 0$. Thus, it follows from the Friedmann
equation
\begin{equation}
\frac{\ddot a}{a}=-\frac{4\pi G}{3}\left(\rho+3p
\right)+\frac{\Lambda}{3} \label{2}
\end{equation}
that  the curve for the scale factor $a$ is always concave
upwards, when drawn as function of cosmic time $t$ (it is here
assumed that $\Lambda \geq 0$). The inequality $\rho+3p <0$ breaks
the  so-called strong energy condition.

Astrophysical data make it quite possible that the value of $w$ is
somewhat less than -1.  It was not understood until rather
recently  that the universe will then have the strange property of
passing into a  Big Rip singularity in the future
\cite{caldwell03,mcinnes02,barrow04}. In turn, this can lead to
bizarre consequences such as negative entropy \cite{brevik04}. As
shown in a recent paper of Nojiri {\it et al.} \cite{nojiri05} one
can now actually  classify as many as four different types of the
Big Rip phenomenon. It should also be noted that quantum effects
which are important near Big Rip may actually act against its
occurrence \cite{nojiri04}.

After providing these introductory remarks we have now come to the
theme of the present work. First, we shall endow the cosmic fluid
with a {\it bulk viscosity} $\zeta$. Such a viscosity is obviously
compatible with spatial isotropy of the universe. In a recent
paper \cite{brevik05} we showed how the physically natural
assumption of letting $\zeta$ be proportional to the scalar
expansion in a spatially flat FRW universe can drive the fluid
into the phantom region even if it starts from the quintessence
region in the non-viscous case. Next, we shall investigate if this
nice property of the fluid still persists if we replace Einstein's
gravity with a more complicated theory. Specifically, we shall
assume the following modified gravity model:
\begin{equation}
S=\frac{1}{16\pi G}\int d^4x\,\sqrt{-g}\left(f_0R^\alpha
+L_m\right). \label{3}
\end{equation}
Here $f_0$ and $\alpha$ are constants, and $L_m$ is the matter
Lagrangian. This is the model recently studied by Abdalla {\it et
al.} \cite{abdalla05}. The dimension of $f_0$ is
$\rm{cm}^{2(\alpha-1)}$.

Our question thus is: does there exist a  crossing of the $w=-1$
barrier, from $w>-1$ to $w<-1$, when the field equations are
derived from the action (\ref{3})? The answer turns out to be yes,
if $\zeta$ satisfies the condition given by Eq.~(\ref{38}) below.
This is actually a quite natural generalization of the condition
found earlier for the case of Einstein's gravity \cite{brevik05}.

The present paper is a continuation of a recent study of viscous
cosmology in modified gravity \cite{brevik05a}. Cf. also the very
recent paper in \cite{ren06} on cosmological models with viscous
media.

We close this section by mentioning the following points. As
discussed in \cite{nojiri05a} and \cite{capozziello05}, a dark
fluid with a time dependent bulk viscosity can be considered as a
fluid having an inhomogeneous equation of state. Also, the
possibility of crossing the $w=-1$ barrier was considered in these
references. Another point is that string/M theory may predict a
modified gravity action with a negative power of $R$
\cite{nojiri03}. A consistent modified gravity theory which
describes not only the cosmic acceleration but may comply with the
Solar System tests was formulated in \cite{nojiri03a}.

\section{The fundamental formalism}

We assume the spatially flat FRW metric,
 \begin{equation}
  ds^2= -dt^2+a^2(t)\,d{\bf x}^2, \label{4}
  \end{equation}
 and put $\Lambda=0$. The four-velocity of the cosmic fluid is
 $U^\mu=(U^0, U^i)$. In comoving coordinates, $U^0=1,\,U^i=0$. In
 terms of the projection tensor
 $h_{\mu\nu}=g_{\mu\nu}+U_\mu U_\nu$ we can write the
 energy-momentum tensor as
 \begin{equation}
 T_{\mu\nu}=\rho U_\mu U_\nu+(p-\zeta \theta)h_{\mu\nu}, \label{5}
 \end{equation}
 assuming constant temperature as well as vanishing shear
 viscosity in the fluid. Here the scalar expansion is
 ${\theta^\mu}_\mu =3\dot{a}/a \equiv 3H$, where $H$ is the Hubble
 parameter. The effective pressure is defined as
 \begin{equation}
 \tilde{p}=p-3H \zeta. \label{6}
 \end{equation}
 From variation of the action (\ref{3}) we obtain the equations of
 motion \cite{abdalla05,brevik05a}
 \begin{eqnarray}
 -\frac{1}{2}f_0\,g_{\mu\nu}R^\alpha+\alpha
f_0\,R_{\mu\nu}R^{\alpha-1}-\alpha f_0\,\nabla_\mu\nabla_\nu
R^{\alpha-1} \nonumber \\
+\alpha f_0\,g_{\mu\nu}\nabla^2 R^{\alpha-1}=8\pi G T_{\mu\nu},
\label{7}
\end{eqnarray}
where $T_{\mu\nu}$ corresponds to the term $L_m$ in (\ref{3}). The
notation is such that the values $\alpha=1,\, f_0 =1$ correspond
to Einstein's gravity.

We first consider the (00)-component of this equation. We
calculate
\begin{equation}
R_{00}=-\frac{3\ddot{a}}{a},\quad
R=6\left(\frac{\ddot{a}}{a}+\frac{\dot{a}^2}{a^2}\right)
=6(\dot{H}+2H^2),\label{8}
\end{equation}
as well as the second order derivatives of $R^{\alpha-1}$. As
$T_{00}=\rho$ we get after some algebra
\begin{equation}
\frac{1}{2}f_0 R^\alpha-
 3\alpha f_0(\dot{H}+H^2)R^{\alpha-1}+3\alpha (\alpha-1)f_0 H
 R^{\alpha-2}\dot{R}=8\pi G\rho \label{9}
\end{equation}
(note that there is a printing error in a sign in the
corresponding (00)-equation (16) in \cite{brevik05a}).

Similarly, we consider the $(rr)$-component of Eq.~(\ref{7}).
Since in a coordinate basis
\begin{equation}
R_{rr}=(\dot{H}+3H^2)a^2, \quad
T_{rr}=\tilde{p}g_{rr}=\tilde{p}a^2, \label{10}
\end{equation}
we get
\[
\frac{1}{2}f_0R^\alpha-\alpha f_0
(\dot{H}+3H^2)R^{\alpha-1}+\alpha (\alpha-1)f_0 \big[
2HR^{\alpha-2} \dot{R} \]
\begin{equation}
+(\alpha-2)R^{\alpha-3}\dot{R}^2+R^{\alpha-2}\ddot{R}\big]=-8\pi
G\tilde{p}. \label{11}
\end{equation}
We consider next the local conservation equation for energy. An
important property of Eq.~(\ref{7}) is the following: taking the
covariant divergence of the expression on the left hand side one
obtains {\it zero}. This was shown explicitly by Koivisto
\cite{koivisto05}. Consequently, we obtain the energy-momentum
conservation equation in conventional form:
\begin{equation}
\nabla^\nu T_{\mu\nu}=0. \label{12}
\end{equation}
Energy-momentum conservation thus occurs as a consequence of the
{\it field equations}, just as in Einstein's theory. This property
strongly supports the consistency of the modified gravity theory.
The
 conservation equation for energy is now obtained by contracting
Eq.~(\ref{12}) with $U^\mu$:
\begin{equation}
 \dot{\rho}+(\rho +p)3H=9\zeta H^2, \label{13}
 \end{equation}
 which can alternatively be written as
 \begin{equation}
 \dot{\rho}=-3\gamma \rho H+9\zeta H^2. \label{14}
 \end{equation}

Let us next search for a differential equation for $H=H(t)$,
wherein $\{f_0, \alpha, \gamma\}$ are given constant input
parameters, and where the bulk viscosity $\zeta(t)$ is, to begin
with, taken to be an arbitrary function of time. The natural way
of accomplishing this task is to differentiate the left hand side
of Eq.~(\ref{9}) with respect to time, thereafter insert for
$\dot{\rho}$ on the right hand side the expression in
Eq.~(\ref{14}), and finally insert for $\rho$ again from Eq.~(9).
We obtain in this way
\[ \frac{3}{2}\gamma f_0HR^\alpha +3\alpha f_0H[2\dot{H}-3\gamma
(\dot{H}+H^2)]R^{\alpha-1} \]
\begin{equation}
+3\alpha
(\alpha-1)f_0H[(3\gamma-1)H\dot{R}+\ddot{R}]R^{\alpha-2}+3\alpha(\alpha-1)(\alpha-2)f_0
H\dot{R}^2R^{\alpha-3}=72\pi G\zeta H^2. \label{15}
\end{equation}
As $R=6(\dot{H}+2H^2)$, this can be regarded as  a nonlinear
differential equation for $H(t)$. As the equation is complicated,
its mathematical structure and physical meaning are best discussed
in terms of examples. In the next section we will discuss the
simplest alternative, $f_0 =1,\, \alpha =1$.

\section{The case $f_0 =1,\, \alpha =1$}

This case is Einstein's gravity. From Eqs.~(\ref{9}), (\ref{11})
and (\ref{15}) we obtain respectively
\begin{equation}
3H^2=8\pi G\rho, \label{16}
\end{equation}
\begin{equation}
2\dot{H}+3H^2=-8\pi G\tilde{p}, \label{17}
\end{equation}
\begin{equation}
2\dot{H}+3\gamma H^2=24\pi G\zeta H, \label{18}
\end{equation}
in accordance with known results (cf., for instance,
Ref.~\cite{brevik94}).

Of particular interest, as shown in \cite{brevik05}, is to take
 $\zeta$  to be proportional to the scalar
expansion through a constant, here called  $\tau_1$. Thus
\begin{equation}
\zeta=\tau_1 \theta = 3\tau_1 H, \label{19}
\end{equation}
On thermodynamical grounds, $\zeta$ has to be a positive quantity.
For an expanding universe, therefore,  $\tau_1$ is also positive.
If the following condition is satisfied,
\begin{equation}
\chi \equiv -\gamma +24\pi G\tau_1 >0, \label{20}
\end{equation}
the equations of motion lead to the occurrence of a future Big Rip
singularity in a finite time $t$. Thus even if we start from an
initial situation where the fluid is non-viscous and being in the
quintessence region ($\gamma >0$), the imposition of a
sufficiently large bulk viscosity will will drive the fluid into
the phantom region and thereafter inevitably into the Big Rip.

On basis of the ansatz (\ref{19}) the equations of motion are
easily solvable as functions of $t$. We take the initial time as
$t=0$, and give a subscript zero to quantities referring to this
instant. Defining the time-dependent quantity $X=X(t)$ as
\begin{equation}
X=1-\chi t \sqrt{6\pi G\rho_0}, \label{21}
\end{equation}
we can express the solutions as
\begin{equation}
H=\sqrt{\frac{8\pi G}{3}\rho_0}\, X^{-1}, \label{22}
\end{equation}
\begin{equation}
\rho=\rho_0\,X^{-2}, \label{23}
\end{equation}
\begin{equation}
p=p_0\,X^{-2}, \quad {\rm{with}} \quad p_0=w\rho_0, \label{24}
\end{equation}
\begin{equation}
R=24\pi G\left( \chi+\frac{4}{3}\right)\rho_0\,X^{-2}, \label{25}
\end{equation}
showing explicitly how the singularities occur when $X \rightarrow
0$.

\section{The case $\alpha =2$}

We choose this case as the next step in complexity. The choice is
motivated chiefly by mathematical convenience. It corresponds to a
gravitational Lagrangian density proportional to $R^2$ in the
action (\ref{3}).

Of main interest is now Eq.~(\ref{15}), in order to make the
appropriate determination of the time dependence of the bulk
viscosity. The equation becomes
\[ \frac{3}{2}\gamma f_0 R^2-6f_0[(3\gamma -2)\dot{H}+3\gamma H^2]R
\]
\begin{equation}
+6f_0[(3\gamma-1)H\dot{R}+\ddot{R}] = 72\pi G \zeta H. \label{26}
\end{equation}
In analogy with Eq.~(\ref{22}) we make the following ansatz for
$H(t)$:
\begin{equation}
H=\frac{H_0}{1-BH_0 t}, \label{27}
 \end{equation}
 where $B$ is a
nondimensional constant. In order for a Big Rip to occur, $B$ has
to be positive. Let us now take  $\zeta(t)$ to be proportional to
the cube of the scalar expansion. Thus, denoting the
proportionality constant by $\tau_2$, we put
\begin{equation}
\zeta =\tau_2 \theta^3=27\tau_2 H^3. \label{28}
\end{equation}
The important point is that by inserting these expressions into
Eq.~({26}) we find the time-dependent factors to drop out. What
remains is an algebraic equation determining the value of $B$ in
terms of the given initial conditions:
\begin{equation}
B^3+\left( 2+\frac{3}{4}\gamma \right) B^2+\frac{3}{2}\gamma B
-\frac{9\pi G\tau_2}{f_0} =0 \label{29}
\end{equation}
(note that the dimension of $f_0$ in this case is
$\rm{cm}^{2(\alpha-1)}={\rm cm}^2$).

The general analysis of this cubic equation leads to rather
unwieldy expressions. The structure of the equation is most
conveniently discussed by means of examples. We define for
convenience the quantity
\begin{equation}
K=\frac{9\pi G\tau_2}{f_0}. \label{30}
\end{equation}

\subsection  {Non-viscous fluid}

  We consider this case for
checking purposes. We now have $\tau_2=0$. From Eq.~(\ref{29}) we
obtain, when the trivial solution $B=0$ is disregarded,
\begin{equation}
B^2+\left(2+\frac{3}{4}\right)B+\frac{3}{2}\gamma=0 \label{31}
\end{equation}
for any value of the constant $\gamma$. Thus both for a "normal"
fluid ($\gamma \geq 1$) and for a quintessence fluid ($0<\gamma
<2/3$) the two solutions of Eq.~(\ref{31}) are both negative (note
that the product of the roots is equal to $3\gamma/2$). There is
thus no Big Rip solution in the non-viscous case when $\gamma \geq
0$, as we would expect.

\subsection {Inclusion of viscosity, $\tau_2 >0$. Vacuum fluid}

 Consider first
the vacuum fluid case, $\gamma=0$. Equation (\ref{29}) yields
\begin{equation}
B^3+2B^2-K =0. \label{32}
\end{equation}
If the left hand side of this expression is drawn as a function of
$B$, we see that there is a local maximum at $B=-4/3$ and a local
negative  minimum  at $B=0$, irrespective of the magnitude of $K$.
There exists thus one single positive root of the equation, for
all positive $K$. This root is caused by the viscosity, and leads
to the future singularity. Let us assume that $K$ (or $\tau_2$)
increases from $K=0$ upwards; then we first encounter a parameter
region in which there exist three real roots \cite{crc60}. Let us
assume this region in the following, and introduce an angle $\phi$
in the interval $0<\phi <180^0$ such that
\begin{equation}
\cos \phi= -\left( 1-\frac{27}{16}K \right). \label{33}
\end{equation}
The solutions can then be written in the form
\begin{eqnarray}
B=\left\{ \begin{array}{lll}
-\frac{2}{3}+\frac{4}{3}\cos\frac{\phi}{3}, \\
-\frac{2}{3}+\frac{4}{3}\cos \left(\frac{\phi}{3}+120^0 \right), \\
-\frac{2}{3}+\frac{4}{3}\cos \left(\frac{\phi}{3}+240^0 \right).
\end{array} \label{34}
\right.
\end{eqnarray}
As an example, we choose the value $K=8/27$, corresponding to
$\phi=120^0$. Then, $B=\{ -1.9196, -0.4351, 0.3547 \}$, where the
last positive solution describes the viscosity-generated Big Rip
phenomenon.

\subsection{Inclusion of viscosity, $\tau_2 >0$. The fluid almost a vacuum
fluid}

As measurements show that the parameter $\gamma$ is close to
unity, it is of physical interest to make a perturbative expansion
around $\gamma =0$. Let us consider the formalism to the first
order in $\gamma$, assuming
\begin{equation}
|\gamma| \ll 1. \label{35}
\end{equation}
The basic equation is Eq.~(\ref{29}), as before. Drawing the left
hand side of the equation versus $B$, we see that the rightmost
extremal point is a local minimum, with coordinates
$(-\frac{3}{8}\gamma, -K)$. Therefore, the equation must have one
positive root. Choosing again the value $K=8/27$ as an example, we
can as before define an angle $\phi$ in the region $0<\phi<180^0$;
it is now determined by
\begin{equation}
\cos \phi=-\frac{1}{2}\left(1-\frac{9}{16}\gamma \right).
\label{36}
\end{equation}
The positive root is then found to be
\begin{equation}
B=-\frac{2}{3}\left(1+\frac{3}{8}\gamma
\right)+\frac{4}{3}\left(1-\frac{3}{16}\gamma \right)\cos
\frac{\phi}{3}, \label{37}
\end{equation}
leading to the viscosity-generated Big Rip in this particular
case. If $\gamma=0$, the first member of Eq.~(\ref{34}) is
recovered.

\section{Remarks on the general case}

For arbitrary values of $\alpha$ we can no longer give complete
solutions, but it turns out that the most important part of the
problem, namely to determine which form of viscosity leads to a
Big Rip, can easily be dealt with. First, we assume the same form
of $H=H(t)$ as before; cf. Eq.~(\ref{27}), where the value of the
parameter $B$ depends on $\alpha$. This means that
$R=6(\dot{H}+2H^2)$ has the same form as before. In analogy with
Eqs.~(\ref{19}) and (\ref{28}) we assume next that $\zeta$ is
proportional to the scalar expansion raised to the power
$(2\alpha-1)$. Denoting the proportionality constant by
$\tau_\alpha$, our basic ansatz thus reads
\begin{equation}
\zeta =\tau_\alpha \,\theta^{2\alpha-1}= \tau_\alpha
(3H)^{2\alpha-1}. \label{38}
\end{equation}
Upon insertion of the expressions for $H, R$ and $\zeta$ into
Eq.~(\ref{15}) we see that the time-dependent factors again drop
out, and we remain with the following equation determining the
value of the constant $B$:
\[ (B+2)^{\alpha-1}\Big\{ 9(2-\alpha
)\gamma+3[\alpha+3\gamma+\alpha(2\alpha-3)(3\gamma-1)]B \]
\begin{equation}
+6\alpha (\alpha-1)(2\alpha-1)B^2 \Big\}
-\frac{144}{f_0}\left(\frac{3}{2}\right)^\alpha \pi
G\tau_\alpha=0. \label{39}
\end{equation}
Because of its complexity this equation has to be analyzed in each
specific case. Any positive root for $B$ leads to a Big Rip
singularity, as follows from the form (\ref{27}) for $H(t)$.

If $\alpha =2$, Eq.~(\ref{39}) reduces to our previous
Eq.~(\ref{29}). It is of interest to reconsider in more detail the
case $\alpha =1$: then Eq.~(\ref{39}) yields the solution
\begin{equation}
B=-\frac{3}{2}\gamma +36\pi G\tau_1. \label{40}
\end{equation}
Since in this case
\begin{equation}
H_0=\sqrt{\frac{8\pi G}{3}\rho_0} \label{41}
\end{equation}
according to Eq.~(\ref{22}), it follows upon insertion that
\begin{equation}
X=1-BH_0t=1-\chi t\sqrt{6\pi G\rho_0}, \label{42}
\end{equation}
where $\chi$ is defined by Eq.~(\ref{20}). This is agreement with
our previous expression (\ref{21}) for $X$ when $\alpha =1$.

\section{Summary}

Our treatment was based on the form (\ref{3}) for the action. Then
assuming the cosmic fluid to possess a bulk viscosity $\zeta$
varying with the scalar expansion as in Eq.~(\ref{38}), we found
how the fluid could in principle be driven into the Big Rip
singularity, even if it stayed in the quintessence region in the
non-viscous case. This is a property following directly from
Eq.~(\ref{27}) for $H$, whenever the constant $B$ is positive. In
general, $B$ is determined by Eq.~(\ref{39}). We worked out the
solutions in reasonable detail when $\alpha=1$ (Einstein's
gravity), and also when $\alpha=2$, assuming a vacuum fluid
($\gamma  \equiv 1+w=0$) as well as almost a vacuum fluid
($|\gamma| \ll 1$).

Finally, we mention that it would be of interest to incorporate
the above formalism in the modified Gauss-Bonnet gravity
interpreted as dark energy \cite{nojiri05b,nojiri05c}.

\section*{Acknowledgment}

It is a pleasure to thank Sergei D. Odintsov for his many valuable
comments on the manuscript.

\end{document}